\documentclass[aps,twocolumn,amsmath,amssymb,showpacs]{revtex4}
\usepackage{epsf}
\begin{document}
\title{The neutron resonance in the cuprates and its effect on fermionic 
excitations}
\author{Ar. Abanov$^{1}$, A. V. Chubukov$^{2}$, M. Eschrig$^{3}$,
M. R. Norman$^4$, and J. Schmalian$^{5}$ }
\affiliation{$^1$Theory Division, Los Alamos National Laboratory, MS B262, 
Los Alamos, NM 87545}
\affiliation{$^2$Department of Physics, University of Wisconsin, Madison, WI
 53706}
\affiliation{$^3$Institut f{\"u}r theoretische Festk{\"o}rperphysik,
Universit{\"a}t Karlsruhe, 76128 Karlsruhe, Germany}
\affiliation{$^4$Materials Science  Division, Argonne National Laboratory, 
Argonne, IL 60439}
\affiliation{$^5$Department of Physics, Iowa State University, Ames, IA 50011}
\date{\today}   
\begin{abstract}
We argue that the exciton scenario for the magnetic resonance in the
cuprate superconductors yields a small 
spectral weight of the resonance, in agreement with experiment. 
We show that the 
small weight is related to its concentration in a small region of
momentum and energy.
Despite this, we find that a large fermionic self-energy can indeed be
generated by a resonance with such properties, i.e., the 
scattering  from the resonance
substantially affects the electronic properties of the cuprates below $T_c$.
\end{abstract}
\maketitle

The magnetic resonance observed in inelastic neutron scattering experiments
in both two-layer (YBCO\cite{YBCO,YBCOx} and Bi2212\cite{BSCO}) and single-layer
(Tl2201\cite{TBCO}) cuprates is one of the most striking 
features of high $T_c$ superconductors. For doping concentrations, $x$,
greater than optimal doping, a sharp peak
emerges at $T_c$ and is resolution limited in energy at low $T$. For $x$ less
than optimal doping, a broadened version appears at a pseudogap temperature
$T^*$, and then narrows in energy  below $T_c$. 
The energy of the peak, $\Omega_{res}$, is found to scale with $T_c$ for all
dopings. The peak is centered at momentum ${\bf Q}=(\pi,\pi)$, and is part of a
collective mode dispersion, with weaker incommensurate ``side branches''
extending to lower energies\cite{DISP}. 

One of the main issues related to the resonance is whether it can account
for the measured changes in the fermionic properties of the
cuprates below $T_c$, via a feedback effect similar to the 
Holstein effect in phonon mediated superconductors. It is not obvious 
that this effect is strong, since 
the total experimental spectral weight of the resonance peak, $I_0 = \int 
S({\bf q},\Omega) d^2q d \Omega/(8\pi^3$), is only a few percent of the 
local moment sum rule\cite{LOCAL}, $S(S+1)/3 = 1/4$. 

In this paper, we address the issue whether the smallness of the
integrated intensity of the peak precludes strong effects on the fermions. Our 
main result is that the fermionic self-energy due to scattering 
from the resonance is strong and unrelated to the small 
integrated intensity of the peak. 
We also discuss the relation between the
resonance peak and the condensation energy

The origin of the resonance has been the subject of 
intense debate in recent years. 
Most theories find that the resonance is a spin exciton
that does not exist in the normal state, but 
emerges in the superconducting state (or, more accurately, when electrons 
acquire a gap) due to a feedback from the pairing on spin collective 
excitations. This effect is specific to $d_{x^2-y^2}$ superconductors
and has no analog for $s-$wave superconductors\cite{KEIMER}.
Moreover, by kinematic constraints, 
the peak at momentum ${\bf Q}$ is due to fermions in the near
vicinity of the ``hot spots'' on the Fermi surface 
(points separated by ${\bf Q}$).
For a  $d_{x^2-y^2}$ superconductor with a Fermi surface like that observed
experimentally, these fermions have a large gap, $\Delta$. As a result, 
spin collective excitations have no damping at $T=0$ up to an energy
$2\Delta$. As the spin exciton is pulled below this $2\Delta$
continuum,  it has a zero linewidth unless other effects, such as impurity 
scattering\cite{SUBIR} and (for $T_c < T < T^*$) the pseudogap, are
incorporated. The displacement 
of $\Omega_{res}$ to lower energies 
from $2\Delta$ increases with underdoping
as the spin-fermion coupling gets larger.

By analogy with the  Holstein effect,  
the emergence of the resonance in the exciton scenario should affect       
the electronic properties of the cuprates: it can give rise to the 
peak-dip-hump features in the fermionic spectral function, most prominent near
the $(\pi,0)$ points\cite{ARPES} where the peak-dip separation equals the
resonance energy\cite{ND,AC}.
It also yields  the kink in the quasiparticle dispersion
along the $(\pi,\pi)$ direction\cite{KINK}, 
with the kink energy 
near $\Delta + \Omega_{res}$\cite{ME}, the
dip in the density of states at about the same energy, and the 
dip in the SIS tunneling conductance \cite{JOHN} and in the 
optical conductivity \cite{OPTICS} at $2\Delta + \Omega_{res}$~\cite{sis,opt}.
It can also cause subgap peaks in SNS
junctions~\cite{assa}.
The issue we address here is the strength of these effects.

 From an experimental perspective,
the features that could be interpreted as due to 
scattering from the resonance have been observed in angle resolved 
photoemission (ARPES) spectra, SIN and SIS tunneling spectra, and
optical conductivity measurements on Bi2212 at various doping 
concentrations~\cite{finger}. Furthermore, the resonance energies 
inferred from ARPES\cite{PRL99} and SIS\cite{JOHN} measurements as a
function of doping match $\Omega_{res}$ as measured directly by neutron
scattering.  The mode extracted from SIS experiments \cite{JOHN} 
is located very near $2\Delta$ in overdoped materials,
but progressively deviates to lower energies with 
underdoping, as would be expected of a collective 
excitation inside a continuum gap~\cite{JOHN}.
In addition, the real part of the fermionic self-energy at the node as a
function of temperature has been shown to scale with the resonance
intensity\cite{KINK}.  It has been claimed, however, that 
other effects such as bilayer splitting, particularly in the overdoped 
cuprates\cite{BILAYER}, and scattering from
phonons\cite{LANZARA}, can also account for these data.

There are two key points to address in the analysis of the 
feedback effect on fermions: the values of the spin-fermion coupling $g$
 and the dimensionless coupling constant $\lambda$,
and the dependence of the self-energy on the integrated intensity of the peak. 

Consider first the issue of the spin-fermion coupling, $g$.
It is defined via the spin-fermion model:
\begin{equation}
H = H_{ferm} + H_{spin} + g S \psi^{\dagger} \sigma \psi 
\end{equation} 
The most straightforward way to extract 
$g$ is to fit the position of the maximum of the 
spin susceptibility 
$\chi^{\prime \prime} ({\bf Q}, \Omega)$ 
in the normal state. 
Experimentally, this maximum is located 
at $20-25 {\rm meV}$ in optimally doped YBCO~\cite{YBCOx}.
The data are consistent with a relaxational form for the susceptibility,
$\chi^{-1} ({\bf Q}, \Omega) = \chi_Q^{-1}
- i \Gamma (\Omega)  $,
whose imaginary part has a maximum given by 
$\Gamma (\Omega_{max}) =  \chi_Q^{-1}$.
Here $\Gamma(\Omega)$ is the imaginary part of the fermionic bubble times
$g^2$, which can be most easily seen by considering the
fermionic bubble as a self-energy insertion in the bosonic (spin fluctuation)
propagator (an equivalent expression is obtained in the random phase
approximation).  The fermionic
bubble  is easily calculated by linearizing the dispersion about the hot
spots ($\epsilon_{\bf k} = v_x k_x + v_y k_y$,
$\epsilon_{{\bf k}+{\bf Q}} = v_x k_x - v_y k_y$) and summing over all 8 hot
spots.  The result is \cite{AC,review}
\begin{equation}
\Gamma (\Omega)  = 2 g^2 \Omega /(\pi v_x v_y)
\label{damping}
\end{equation} 
At the hot spots, $v_x \approx v_y \approx v_F/\sqrt{2}$, 
where $v_F$ is the Fermi velocity at the hot spots. 
Using the  experimental $\chi_{Q} \sim 13 {\rm states}/{\rm eV}$~\cite{YBCO1},
$\Omega_{max}=20 {\rm meV}$,
and $v_F \sim 0.4 {\rm eV}$~\cite{comm}
(in units where the lattice constant is 1) 
we then obtain $g \sim 1.75 v_F \sim 0.7 {\rm eV}$.

The dimensionless coupling $\lambda$ can be extracted from 
the fermionic  self-energy at the lowest $\omega$:
$Re\Sigma (\omega) = -\lambda \omega$. At the same level of approximation
as Eq.~2, $\Sigma ({\bf k},\omega)$ is determined as 
3$g^2$ times a convolution of $\chi({\bf q},\Omega)$ with 
$G_0({\bf k}+{\bf q},\omega+\Omega)$
($G_0$ is the fermion Green's function, and the factor of 3 is due to spin
summation).  Again, linearizing the fermionic dispersion about the hot
spots, and expanding $\chi$ quadratically about ${\bf Q}$
with a correlation length, $\xi$, we obtain \cite{review}  
\begin{equation}
\lambda = 3 g^2 \chi_{Q}/(4 \pi v_F \xi) = 3 v_F/(16 \Omega_{max} \xi)
\end{equation}
Substituting the above numbers and $\xi \sim 2$, we find $\lambda \sim 2$.
We note that $\lambda$ by definition refers to fermions near the hot spots, 
and is obtained by coupling to the entire spin fluctuation spectrum.

Our value of $g$ is consistent with
fitting resistivity data to spin fluctuation 
scattering\cite{MP} and with Eliashberg calculations of $\Delta$ and
$\Omega_{res}$~\cite{finger}.  Such a large value of $g$ is also expected on 
microscopic grounds:  in the Hubbard model, the effective $g$
is expected to be of the order of the fermionic bandwidth $W$~\cite{VT} 
which is $1 {\rm eV}$ for the cuprates.
Our estimate  $\lambda \sim 2$ is consistent with the
velocity renormalization estimated from normal state ARPES 
experiments\cite{OLSON}. 
Moreover, the specific heat: $C = \gamma T$ with 
experimental 
$\gamma \simeq 2\frac{\rm mJ}{{\rm g-at}\text{ } K^2}$ (Ref.~\cite{loram})
 yields in a two-layer system $N_{0}\sim 
 \frac{2.8}{1+\lambda} {\rm eV}^{-1} \sim 1 {\rm eV}^{-1}$,
where $N_0$ is the (bare) density of states per spin. 
 Again, 
this $N_0$ is close to our value $N_0 \sim 1/(\pi v_F) \sim 1  {\rm eV}^{-1}$
 (using the previously mentioned number for $v_F$).

The result that  $\lambda > 1$ might question the validity
of Eqs.~2 and 3. 
In general, a Migdal theorem does not exist for spin fluctuations,
since spin fluctuations are made out of fermions, and hence
the bosonic energy scale is comparable to the (renormalized)
Fermi energy for general ${\bf q}$.  Our theory, though, is based
on an expansion of fermionic degrees of freedom about the hot
spots, and bosonic degress of freedom about ${\bf Q}$.  That is,
high energy excitation processes have been integrated out, and
are absorbed into the definition of $\chi_Q$.  This means that
in the context of our theory, only low energy vertex corrections
are relevant, and they are unimportant
for the same reason as in the
electron-phonon problem, that is spin fluctuations are slow
compared to fermions~\cite{review}.  
For this reason, an ``effective'' Migdal theorem exists, and
justifies Eqs.~2 and 3.

We next discuss the spectral weight of the resonance, and how this
affects the fermionic self-energy in the superconducting state. 
We begin by noting that since the resonance peak is strong 
at ${\bf Q}$, if it was present for all momenta, the 
total integrated intensity would be $O(1)$. However, the peak 
only exists in a momentum range between ${\bf Q}$ and ${\bf Q}_{min}$, where
${\bf Q}_{min}$ is 
the momentum connecting the nodal points at the Fermi surface.
This occurs since the particle-hole continuum extends to zero frequency
at ${\bf Q}_{min}$, and the resonance ceases to exist there. 
As ${\bf q}$ approaches ${\bf Q}_{min}$, both
the energy and the intensity of the resonance peak vanish
(that is, the incommensurate side branches).
This behavior is consistent with the observed ``negative'' 
curvature of the resonance dispersion and the progressive
reduction of the peak intensity as ${\bf q}$ deviates from ${\bf Q}$
~\cite{neg_disp}.  The ARPES measurements
of the Fermi surface all show that near optimal doping, 
${\bf Q} -{\bf Q}_{min} 
\approx 0.2 (\pi, \pi)$, i.e., the resonance peak 
exists in a momentum range which constitutes only $6\%$ of the area of 
the Brillouin zone. This smallness of the momentum range 
 gives rise to the smallness of $I_0$.

The real issue, though, is whether a small $I_0$
implies a small fermionic self-energy.
We argue that it does not.
As just stated, the resonance peak is strong, but exists only in a
limited momentum range, which is why $I_0$ is small.  Whether or not the
small momentum integrated intensity of the peak matters then depends  
on whether or not typical bosonic momenta that contribute to the fermionic
$\Sigma ({\bf k}, \omega)$ are within the allowed ${\bf q}$ range of the peak.
If they are not, then the smallness of $I_0$ matters.
If they are, then it does not.  These
typical ${\bf q}$ can be easily estimated from an analysis of the
fermionic self-energy ($\Sigma \sim \int G_0 \chi$) discussed above, and 
for $\omega \leq 100 {\rm meV}$ at which the resonance mode affects the 
self-energy, are well within the 
range between ${\bf Q}$ and ${\bf Q}_{min}$ for fermions near the hot
spots: $|{\bf Q}-{\bf q}|\sim\omega/v_F\approx 0.08\pi\ll \sqrt{2} (0.2\pi)$. 
Thus, although the resonance peak  
occupies only a small portion of the
Brillouin zone, it is actually broader than the typical
momentum scale for fermions. 
In this situation, in the
calculations of the fermionic self-energy, one can approximate  
$S({\bf q}, \Omega)$ by its large  
value  in the near vicinity of ${\bf Q}$, 
and neglect the dispersion of the peak. This in turn implies that the 
small $I_0$ {\it does not} matter for the self-energy.
We note again in this regard that experimentally~\cite{YBCO1} 
$\int d \Omega S({\bf Q}, \Omega) \sim 1.6$ is indeed not small. 

We next 
discuss the relation between the resonance peak and the 
condensation energy. The issue is whether the resonance peak contribution to the
condensation energy is consistent with experiment. 
This issue is somewhat
non-trivial as the internal energy is the sum of the kinetic and the
potential energies. The exchange part of the potential energy is related to
the difference between the integrated $S({\bf q},\Omega )$ in the normal and
the superconducting states\cite{SW}: 
$E_{p} =(3J/16\pi^3) \sum_{i} \int d^{2}qd\Omega~(S^{(i)}_{n}({\bf q},\Omega )
 -S^{(i)}_{sc}({\bf q},\Omega ))(\cos q_{x}+\cos q_{y})$
(the summation over $i=o,e$ goes over odd and even channels 
in two-layer systems). If we
assume that $S^{(e)}$ and $S^{(o)}_{n}$ are negligible, 
and approximate $S^{(o)}_{sc}({\bf q},\Omega )$
by $S^{(o)}_{sc}({\bf q},\Omega )=
\pi \chi _{Q}\Omega _{res}\delta (\Omega -\Omega_{res})$ for 
$|{\bf Q}-{\bf q}|<|{\bf Q}-{\bf Q}_{min}|$, and 
$S^{(o)}_{sc}({\bf q},\Omega )=0$
elsewhere, we find 
$E_{p}= 3J(2\pi \chi_{Q}\Omega_{res})(|{\bf Q}-{\bf Q}_{min}|/4\pi )^{2}
\approx 0.05J$. Similar values are
found from explicit calculations using the random phase approximation\cite
{NORM00}. This energy savings is already a small number. The actual value of
the potential energy is, however, even smaller due to compensation from 
$S_{n}$ and other non-exchange terms in the Hamiltonian. Eliashberg-type 
computations of the potential energy including the
normal $S_{n}$ part but without taking into account the restriction on 
${\bf q}$ found~\cite{ac_ce} $E_{p}\sim 0.008J\sim 10K$. The restriction
on ${\bf q}$ further reduces this energy.
The condensation energy $E_{c}$
extracted from the specific heat measurements is $E_{c}\sim 10K$.
This condensation energy is about $E_{p}$. In reality, $E_{c}$ should be
greater than $E_{p}$
since at strong coupling, the kinetic energy decreases
in the superconducting state as the fermionic excitations become
less diffusive in the superconducting state due to feedback effects again
associated with the resonance~\cite{COND00}. In any event, we clearly see
that the resonance viewed as a spin exciton for which the intensity at 
${\bf Q}$ is not small yields  a
small value of $E_c$, in agreement with the data.

It is also essential to point out 
an important difference between the coupling 
of fermions to the resonance mode in a $d$-wave
superconductor and the coupling of fermions to 
antiferromagnetic magnons.
In the latter case, the spin mode couples to fermions only through gradients,
i.e., the renormalized coupling $g_{eff}$ is much smaller than $g$. 
This reduction from $g$ to $g_{eff}$ is the result
of strong vertex corrections if antiferromagnetic magnons are present 
in the normal state~\cite{WARD}, and occurs because antiferromagnetic magnons
are only compatible with a small Fermi surface (hole pockets), in which 
case $g$ has been absorbed into the definition of renormalized   
fermions with an SDW energy gap \cite{KAMPF}.
However, we are treating the metallic phase near optimal doping where a large 
Fermi surface exists, the normal state spin dynamics is purely relaxational, 
and the resonance peak appears {\it only} when fermions acquire 
a d-wave superconducting gap.  Thus, $g$ is the appropriate coupling to use,
not $g_{eff}$.  The crossover between these two regimes should occur
in the low doping regime where the
Fermi surface evolves towards small hole pockets. 
(Note in passing that for these reasons, the 
resonance mode is not the ``glue'' for the 
magnetically mediated pairing theory near optimal doping - this pairing is 
produced by {\it overdamped} spin excitations.)

The above picture of the spin resonance and its effect on fermions 
has been challenged by a number of authors.  Perhaps the
work which best summarizes these objections is that of Ref.~\onlinecite{KEE}.
They  argued that $g \sim 14 {\rm meV}$ and $\lambda \sim 10^{-3}$,
two and three orders of magnitude smaller than our values, respectively.
The large difference in $g$ is the combination of several factors.
First, the value of $\Gamma$ that we 
extracted from the data is about 8 times larger than theirs.
This is because they equated Eq.~2 with the half width of the resonance,
without taking into account the fact that the resonance width is
strongly reduced compared to the normal state because of gapping
of the particle-hole continuum.
Moreover, from Eq.~2, we see that the full width of the normal state
(relaxational) $\chi$ is not 2$\Gamma$, but rather
$2\sqrt{3} \Omega_{max}$ where $\Gamma(\Omega_{max}) = \chi_Q^{-1}$.
Second, they assumed $\Omega_{max}$ was the resonance mode energy (40 meV),
whereas we used the normal state maximum (20 meV).
Third, they assumed an $N_{0}\sim J^{-1}\sim 10{\rm eV}^{-1}$,
i.e., their $v_F$ is about 12.5 times smaller than ours. 
The combination of these three factors 
accounts for the factor of 50 difference in $g$.  The even larger 
discrepancy in
$\lambda$ is due to their coupling only to the resonance (which they treat
as an Einstein mode), and not to the 
entire spin fluctuation spectrum as we have done.

Moreover, they  approximated  the
resonance peak as a product of two $\delta$-functions:
$S ({\bf q},\Omega) = (2\pi)^3 I_0 \delta ({\bf q}-{\bf Q}) \delta (\Omega -
\Omega_{res}) + ...$
where dots stand for the non-resonance part.  Using this 
approximation, the estimated fermionic self-energy due to 
spin-fermion scattering scales as $I_0 (g/\Omega_{res})^2$. 
By their estimates, 
$g \sim 0.35 \Omega_{res}$, $I_0 \ll 1$, and hence the effect on 
the fermionic self-energy is negligible.
As demonstrated above, our estimate for $g$: $g \sim 0.7 {\rm eV }
\sim 17 \Omega_{res}$  is very different from theirs.
This is the primary reason we get a large
self-energy, and they get a small one.
We emphasize, however, that the approximation that
$S({\bf q}, \Omega)$ is a $\delta$ function in momentum space leads to 
a qualitatively incorrect fermionic self-energy, in that, as stated earlier,
the typical fermionic momentum scale is actually smaller than the resonance
width.  Furthermore, in such an approximation,
the imaginary part of the self-energy is simply a $\delta$-function in energy,
since only one bosonic momentum contributes.
This is clearly not consistent with experiment.
The aspect which is completely missed by using
the $\delta$-function approximation in ${\bf q}$ is that as soon as the 
resonance 
has a finite width in momentum space, the internal momentum sum in the 
Feynman diagram for the self-energy is dominated by the flat fermionic 
dispersion in the vicinity of the $(\pi,0)$ points\cite{ME}.  This is why the 
self-energy effects are so large for momenta near $(\pi,0)$, and also why 
the energy scale at which structure appears in the spectral function 
(spectral dips and kink energies) is independent of momentum\cite{KINK}.

To summarize, we demonstrated in this paper that the 
large intensity of the resonance at ${\bf Q}= (\pi,\pi)$ is
consistent with the small value of the total momentum and frequency
integrated intensity of the resonance peak, and with the fact that the
magnetic part of the condensation energy is only a small fraction of $J$.
We found that the spin-fermion coupling $g$ is of the order of $1 {\rm eV}$
and argued that this value of $g$ is consistent with experiment.
This $g$ is sufficiently large that scattering from the resonance can
substantially affect the electronic properties of the cuprates below $T_c$.

This work was supported by the U.S. Dept. of Energy,
Office of Science, under contracts W-31-109-ENG-38 (M.R.N. and M.E.) and
W-7405-ENG-82 (J.S.),
NSF DMR-9979749 (A.C.), and DR Project 200153 at LANL (Ar.A.).


\begin{thebibliography}{99}
\bibitem{KEIMER} This occurs 
(i) because of the differences in the coherence 
factors in the BCS susceptibility bubble, and (ii) because the 
$d-$wave pairing involves fermions from different magnetic sublattices and 
does not affect the magnetic correlation length.
\bibitem{comm}  This $v_F =0.4 {\rm eV}$ is about half the unrenormalized 
velocity along the $(\pi,\pi)$ direction. 
This factor of $1/2$ comes from a 
$t-t^\prime$ model fit to the measured Fermi surface near optimal doping.
\end{thebibliography}
\end{document}